\documentclass[11pt, oneside]{article}   	
\usepackage{geometry}      
\usepackage{amsmath}
\usepackage{cancel}
\usepackage{caption}
\usepackage{subcaption}
\usepackage[section]{placeins}

\usepackage{textgreek}

\usepackage[colorlinks=true, allcolors=blue]{hyperref}

\usepackage[usenames, dvipsnames]{color}

\usepackage{amsfonts}
\usepackage{mathrsfs}
\usepackage{adjustbox}
\usepackage{amssymb}
\usepackage{graphicx}				
\usepackage{subcaption}
\usepackage{gensymb}			
\usepackage{amssymb}
\usepackage{float}

\usepackage{lpic}

\usepackage[affil-it]{authblk}

\usepackage[aboveskip=1pt,labelfont=bf,labelsep=period,justification=raggedright,singlelinecheck=off]{caption}

\title{\begin{Huge}
{Threshold Response to Stochasticity in Morphogenesis}
\end{Huge}}
\date{April 4, 2018
}
\author[1]{George Courcoubetis}
\author[2]{Sammi Ali }
\author[2]{Sergey V Nuzhdin}
\author[3]{Paul Marjoram}
\author[1]{Stephan Haas}
\affil[1]{Department of Physics \& Astronomy, University of Southern California, Los Angeles,
California, United States of America}
\affil[2]{Department of Molecular and Computational Biology, University of Southern California, Los Angeles,
California, United States of America}
\affil[3]{ Department of Preventative Medicine, Keck School of Medicine of
USC, Los Angeles, California, United States of America }

\begin{document}
\bigskip
\bigskip
\bigskip
\bigskip
\bigskip
\bigskip

\maketitle

\abstract{During development of biological organisms, multiple complex structures are formed. In many instances, these structures need to exhibit a high degree of order to be functional, although many of their constituents are intrinsically stochastic. Hence, it has been suggested that biological robustness ultimately must rely on complex gene regulatory networks and clean-up mechanisms. Here we explore developmental processes that have evolved inherent robustness against stochasticity. In the context of the Drosophila eye disc, multiple optical units, ommatidia, develop into crystal-like patterns. During the larva-to-pupa stage of metamorphosis, the centers of the ommatidia are specified initially through the diffusion of morphogens, followed by the specification of R8 cells. Establishing the R8 cell is crucial in setting up the geometric, and functional, relationships of cells within an ommatidium and among neighboring ommatidia. Here we study a mathematical model of these spatio-temporal processes in the presence of stochasticity, defining and applying measures that quantify order within the resulting spatial patterns. We observe a universal sigmoidal response to increasing transcriptional noise. Ordered patterns persist up to a threshold noise level in the model parameters. As the noise is further increased past a threshold point of no return, these ordered patterns rapidly become disordered. Such robustness in development allows for the accumulation of genetic variation without any observable changes in phenotype. We argue that the observed sigmoidal dependence introduces robustness allowing for sizable amounts of genetic variation and  transcriptional noise to be tolerated in natural populations without resulting in phenotype variation.


\newpage

\section*{Author summary}

The development of biological organisms  requires the formation of highly ordered structures. These structures are created utilizing cell signaling through morphogens, a process that is inherently stochastic. In this paper, we apply rigorous measures to quantify order and disorder in the context of the development of the Drosophila eye disc, R8 pattern formation in the larva-to-pupa stage. Specifically, we turn to mathematical formulations of the mechanism, and we illustrate that introducing noise below a certain threshold does not affect the final outcome at all. Furthermore, we show that after this threshold is crossed, there is rapid deterioration of the pattern that intensifies as the size of the eye disc is increased. Through quantitative analysis, we explain the origin of both the robustness and rapid deterioration of the pattern. These findings highlight generic characteristics that are needed for developmental models to reproduce experimentally observed collective mechanisms, such as cleanup of superfluous cells and cell threshold responses to stochasticity. We discuss the connections of in these findings with canalization and cryptic genetic variation, i.e. the fact that environmental differences and stochastically perturbed genetic makeups result to identical phenotypes.

\section*{Introduction}

%
%
\subsection*{Deterministic Outcomes from Inherently Stochastic Components}

Biological systems are intrinsically noisy but nonetheless produce deterministic outcomes. During development, organisms utilize signaling molecules, i.e. morphogens, to generate a body plan and differentiate cells. With modern experimental techniques, it is possible to measure temporal concentrations of selected morphogens in each cell. The expression of a gene depends on the probabilistic outcomes of several factors, such as molecular binding affinities, processivity, and regulatory sequence interactions. Specifically,  genetically identical cells produce morphogen transcripts in asynchronous bursts and in varying quantities ~\cite{chubb2006transcriptional}. It is therefore essential to investigate how ordered deterministic structures are formed from underlying stochastic components, such as gene expression. Early studies have addressed robustness of developmental processes, also termed as canalization~\cite{waddington1959canalization}, which remain subject of great interest today~\cite{avishai2017dealingwithnoise, barkai2009robust,gursky2012mechanismsofdevrobust,keller2002developmentalrobustness}. Developmental robustness has been highlighted as a necessary condition that narrows down dramatically the search for plausible models and one that can even ``predict key mechanistic and molecular properties of the associated biochemical circuits"~\cite{avishai2017dealingwithnoise}. For example, the Turing mechanism has been deemed inapplicable in many instances due to its sensitivity to noise common in developmental processes~\cite{keller2002developmentalrobustness}. In this paper, we analyze the response of a developmental system to noise quantitatively from a statistical physics perspective.

\subsection*{Pattern Formation in the Drosophila Eye Disc is an Ideal Model System to Study the Effects of Stochastic Transcription on Deterministic Development}
The compound eye, found primarily in insects, consists of an array of repeating visual units. The Drosophila compound eye is made up of approximately 800 unit eyes, known as ommatidia. In wild-type Drosophila, the structure of the ommatidia resembles a near-perfect hexagonal lattice. This highly structured pattern is developed via the delicate coordination of cell signaling, proliferation, movement and apoptosis \cite{KumarBuild2012}. Some of these cellular processes are guided through the communication of a few conserved molecules, known as morphogens, resulting in tissue morphogenesis. 

While the resulting functional eye emerges in the adult, the role of each cell within repeating ommatidial arrays, and other head structures, are specified during larval development~\cite{KumarBuild2012}. The larval eye-antenna imaginal disc, hereafter referred to as the eye disc, contains numerous cells that produce various morphogens. The Drosophila eye disc is one of the simplest systems for studying morphogenesis since this tissue develops in a mostly  two-dimensional fashion. This is because the eye-antenna disc is derived from an epithelial monolayer~\cite{Haynie}. Furthermore, this system is an ideal one in which to model morphogenesis, since cell lineage has a minimal effect on pattern formation~\cite{Lawrence}, and differentiation relies primarily on cell-to-cell signaling. The underlying developmental mechanisms that guide the formation of the eye disc from larva to adult have been studied extensively, allowing for the generalization and  analysis of  biologically realistic models.

To investigate the stability of the developmental pathways in the Drosophila eye disc, mathematical modeling and numerical simulations are used and the results are analyzed here from a physics perspective.  Mathematical modeling of the Drosophila eye disc pattern-formation mechanism has been the focus of previous investigations,  ~\cite{Barkai,Lubensky} which proposed a mathematical model that reproduces the triangular lattice pattern of differentiated R8 cells and that is robust enough to be biologically plausible. Using this latest model as a basis, we examine and quantitatively test the robustness of the emerging spatially ordered patterns of differentiated R8 cells when transcriptional noise is introduced. To this end, we implement various appropriate measures of spatial order, testing the functional relationship of R8 cell pattern order with increasing stochasticity. 

In this model, morphogens are produced and emitted by individual cells in quantities that are determined by morphogen inputs from other cells in their vicinity. In addition, the production and diffusion  of morphogens is non-uniform, i.e. a given morphogen concentration in a cell does not result in a single production or diffusion rate. These processes are thus inherently noisy. Therefore, the highly ordered structure of the ommatidia in the Drosophila eye disc, emerging from such stochastic constituents, requires robustness in the underlying gene regulatory networks. Our study aims to quantify eye disc developmental robustness. We focus on one specific step out of many involved in eye disc development, in which additional robustness is introduced \cite{Wolffcelldeath}. However, the regulation of R8 cell distances is a crucial developmental step in precisely placing each repeating eye unit, thereby strongly influencing the organism's overall visual acuity.

\subsection*{Mechanism Leading to R8 Cell Specification }

Each ommatidium is made up of 14 cells; 8 photoreceptors (R1-R8), 4 cone cells, and 2 primary pigment cells. The specification of the first photoreceptor, R8, guides the specification and orientation of the remainder of the cells within the ommatidium. Thus, the spacing between neighboring R8 cells is pivotal for positioning and refining geometric relationships in resulting ommatidia. Cellular defects that misplace a single ommatidium will influence the position of neighboring ommatidia, thus propagating further flaws in the lattice. If even a single gene involved in ommatidia formation is perturbed,  all unit eyes can be affected, since the regulatory logic that generates each ommatidium is repeated using the same set of morphogens and gene networks~\cite{KumarBuild2012}.

Prior to R8 cell specification, the eye disc is composed of tightly-packed undifferentiated cells dividing asynchronously. Differentiation starts with the initiation of the morphogenetic furrow (MF), a physical indentation in the eye disc that sweeps through the tissue to dictate the pace at which ommatidia are specified and positioned. The MF advances anteriorly via the communication of several morphogens, primarily Hedgehog (Hh) and Decapentapegic (Dpp). It is initiated in the posterior end of the eye disc and sweeps through the tissue towards the anterior end. As the MF sweeps through the eye disc, the cells posterior to the MF differentiate and commit to their respective role within the ommatidium. Anterior to the MF, the cells are asynchronously proliferating. At the anterior interface of the MF, cells are arrested in G1 to allow synchronous divisions~\cite{wolff1991beginning}. Within the MF, the central R8 photoreceptor cells are specified via the expression of atonal (ato). The spacing between ato-expressing cells is refined posteriorly within the MF. This process starts with 5-7 ato-expressing cells (intermediate groups), followed by 3 ato-expressing cells (equivalence groups) and finally a single ato-expressing cell (R8)~\cite{Sun}. Controlling the spacing of R8 cells within and between columns during MF progression is a crucial step in establishing the spacing pf ommatidia within and between visual columns.

Determining the central R8 cell drives the subsequent specification of the other cells within the ommatidium. Hedgehog (Hh) induces long-range and short-range secondary signals that control the precise position of the MF. Hh acts over a short range to induce the expression of Decapentapegic (Dpp), which in turn diffuses over a long range to turn off homeothorax (hth) and turn on hairy (h), establishing a pre-proneural domain (PPN). Hh and Dpp also induce the expression of Delta (Dl), a trans-membrane ligand that acts on adjacent cells to turn off hairy. DI also acts on the intermediate groups to help refine the ato-expressing cluster, ultimately resulting in a single R8 cell per equivalence group~\cite{Baker}. The single ato-expressing R8 cell produces morphogens that inhibit nearby cells from further expressing \textit{ato}. This ensures regular spacing between R8 cells. The progression of the morphogenetic furrow allows R8 specification to be repeated unidirectionally from posterior to anterior, while the regulation of the distance between neighboring R8 cells generates the remarkable hexagonal array.
%
%

\subsection*{Process Leading to Pattern Formation in the Drosophila Eye Disc}

Here we provide a simplified description of the rational behind the current model, describing the basic mechanism of pattern formation in the eye disc. This process is described using a system of coupled differential equations intended to provide the reader with an intuitive understanding of the foundations of the mathematical model. The details of the full model are elaborated on in the methods section below.

The cells of the eye disc are assumed to lie on a $N\times N$  hexagonal grid, as appropriate for a hard sphere tight packing. After the MF passes, a portion of these cells differentiate to become activated R8 cells, and the remainder will be left undifferentiated. At the end of the process, activated R8 cell centers are positioned on a triangular lattice. This process is  illustrated in Fig~\ref{R8visualiation}.

Some morphogens are diffusible, while others are cell-specific and stay within the boundaries of the cell. In this study, {\textit{ato}} is  the only non-diffusing, cell-specific morphogen, while all others can diffuse between cells. The various morphogens either have inhibitory or activating properties. The inhibitors are morphogens that, when present in a cell, decrease the rate of production of another morphogen, and vice-versa for the activators. 

Initially, differentiated R8 cells are defined as cells that express any non-zero level of \textit{ato} . There are two main ingredients that make the pattern of R8 cells form. First, there is an inhibition signal that blocks atonal. When a cell becomes differentiated, it instantly produces an inhibitory signal that completely blocks the production rate of \textit{ato} of all cells within its vicinity, a fixed circular region. Second, there is an inductor (or activator) signal that causes undifferentiated cells that do not receive the inhibition signal to produce atonal. The activator signal represents the ultimate function of the MF in the mathematical model. It is simply a rectangular wave form, with one edge expanding with a constant velocity. It eventually moves through the entire hexagonal lattice. Cells within this area that are not receiving inhibitory signals will differentiate.  

As an initial condition, a periodic array of differentiated cells are placed as in the posterior region. The spacing of these cell clusters is chosen such that it will allow for the pattern to propagate. In this simplified setting, the initial configuration is defined by differentiated cells separated evenly in a single row in the posterior-most region. The MF then starts to propagate from posterior to anterior, and it activates cells that do not receive the inhibition signal. This causes the pattern to propagate as shown in  Fig~\ref{R8visualiation}.

The actual model that leads to pattern formation is more elaborate. Now that there is a basic understanding of how the pattern propagates, some more intricate aspects can be introduced. In order for the model to be biologically plausible, cell clusters are formed instead of single cells. Even in this illustrative context, as it can be inferred schematically from Fig~\ref{R8visualiation}, the propagation of the pattern is very sensitive to the shape of the inhibitory regions. A recent refinement of the model eliminates this flaw by including a diffusible activator. In simple terms, this forces the first uninhibited cell to receive the linearly propagating activator signal  to create a circular activator region. This activator region defines the cluster size and shape and makes the model robust, preventing the creation of catastrophes, as explained in Ref.~\cite{Barkai}.

\begin{figure}[H]
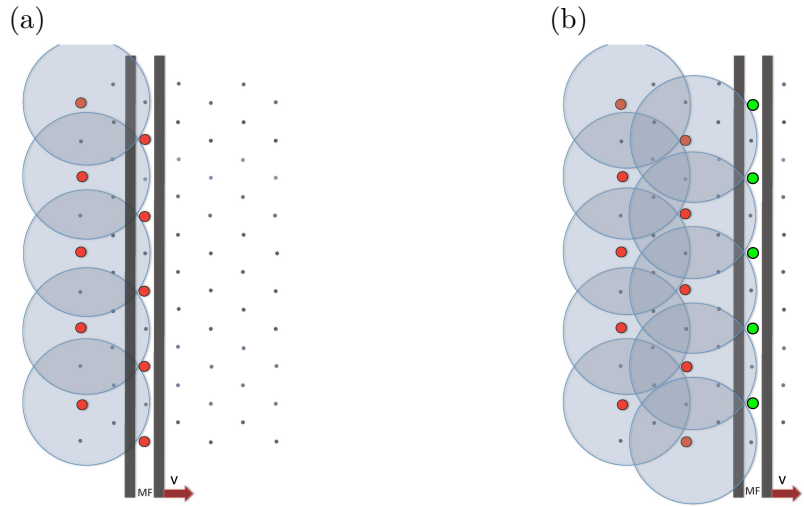

\centering
 \begin{lpic}[l(10mm),r(10mm),t(10mm),b(10mm),grid(100),coords(100),draft,clean]{cartoon(0.06,0.06)}
\lbl[t]{280,1200; (a)}
\lbl[t]{1480,1200; (b)}
\end{lpic}
\caption{ {\bf Visualization of the foundations of the R8 cell specification mechanism in the Drosophila eye disc.}  Simplified illustration of  pattern formation mechanism in the Drosophila eye disc as a result of the competition between short-range inhibitor and long-range activator morphogens. Posterior region of the eye disc with an initial row of differentiated precursor cells, denoted by red dots. The morphogenetic furrow (MF), modeled by a plane wave front, moves to the right towards the anterior region. The gray circles represent the boundaries of regions affected by the short-range inhibitor, where the morphogenetic furrow (MF) will not initiate production of \textit{atonal}. Therefore differentiation can only occur in those locations which are not affected by the short-range inhibitor, leading to the hexagonal supper-lattice of differentiated R8 cells, which is  experimentally observed.  \label{R8visualiation}}  
\end{figure}

\newpage

\section*{Methods}

\subsection*{Mathematical Formulation}

We study models of coupled differential equations that describe morphogenetic pattern formation, relying on the simplifying assumption that the number of cells in the eye disc is fixed~\cite{Barkai,Lubensky}. The ommatidia are arranged in a hexagonal grid, treating it as an underlying two-dimensional structure. Every ommatidium, due to its hexagonal geometry, has six adjacent neighbors. The resulting coupled ordinary differential equations are of the form:

\begin{equation} \label{eq:1} 
\tau_a \frac{da^i}{dt} = P_a \theta(a^i - a_a ) - \lambda_a a^i + G \theta(h^i-h_1)(1-\theta(u^i-u_1) ) + S \theta (s^i-s_1) (1- \theta(u^i-u_1) ),\end{equation}
\begin{equation}  \label{eq:2}
\tau_u \frac{du^i}{dt} = P_u \theta(a^i-a_u) - \lambda_u u^i + D_u \Delta u^i ,\end{equation}
\begin{equation}  \label{eq:3}
\tau_s \frac{ds^i}{dt} = P_s \theta( a^i - a_a ) - \lambda_s s^i + D_s \Delta s^i ,\end{equation}
\begin{equation} \label{eq:4} 
h^i(t)=\begin{cases} P_h\Big(1- (\frac{v\tau_h+c_1}{2c_1})\exp{[\frac{-v\tau_h+c_1}{2D_h}(y- vt)} ]\Big) \, \qquad y\leq vt 
\\ 
P_h(\frac{-v\tau_h+c_1}{2c_1})\exp{[\frac{-v\tau_h-c_1}{2D_h}(y- vt)} ] \qquad y> vt \end{cases}
\end{equation}

These represent the spatio-temporal evolution of the morphogen concentrations \textit{a}, \textit{u}, \textit{s}, and \textit{h}, responsible for pattern formation in the drosophila eye disc. Here, the upper case indexes label each cell. The pro-neural transcription factor marking the center of the future ommatidia is \textit{a} (for \textit{atonal}), \textit{u} represents all diffusible inhibitors, \textit{s} denotes diffusible activator (\textit{scabrous}), and \textit{h} (for \textit{hairy}) describes the morphogenetic furrow (MF) which activates production of atonal along a propagating wave front.  MF propagation is mathematically derived to have this functional form from the underlying morphogen differential equations
\cite{Barkai}. The constant \textit{v} is the velocity of the MF, \textit{D}\textsubscript{h} is the diffusivity, \textit{P}\textsubscript{h} the production rate, and \texttau{}\textsubscript{h} is the reaction time scale of \textit{hairy}.

The mathematical model contains multiple parameters, operators and functions that play important roles in the dynamics of the system. First, the characteristic reaction time scale of each morphogen is given by corresponding \texttau's on the left hand side of the equations. In addition, each morphogen's differential equation contains a term with a $-\lambda$  coefficient that incorporates the spontaneous decay of the morphogens. This term, combined with \texttau, determines the mean lifetime of the morphogen via the ratio $\frac{\tau}{\lambda}$. In addition, for the morphogens that diffuse, there are Laplacian operators associated with diffusion, \textDelta, which are discretized at the cell level (see S1~Appendix for exact expression). For cells on the boundary, reflective boundary conditions are used. Note that although only the equations for \textit{u} and s have this diffusion term, the differential equations that lead to the functional form of \textit{h} are also based on diffusion. Each Laplacian operator is multiplied by a term \textit{D}, the value of which (when divided by the respective \texttau) sets the scale how fast the morphogen diffuses.  

The theta functions are essential, as they set the morphogen concentration thresholds that start, accelerate and stop production. They are a simplified version of the more biologically plausible sigmoid functions. They are defined by
\begin{equation} \label{eq:5} \theta(x)=1 \quad  x>0,\quad \quad  \theta(x)=0 \quad x<0. \end{equation}
The  theta function  $\theta(h^i-h_1)$ in Eq~(\ref{eq:1}) is responsible for initiating production of atonal, \textit{a}, in the target cell with a rate $\frac{G}{\tau_a}$, once the concentration of \textit{h} crosses the threshold \textit{h}\textsubscript{1}. This term is multiplied by $(1-\theta(u^i-u_1))$, a function that goes to zero when $u^i>u^1$. This term introduces the functionality of u as a direct inhibitor of atonal in the model. The last term in Eq~(\ref{eq:1}) introduces \textit{s} as an activator of hairy. Furthermore, the first term in Eq~(\ref{eq:1}), $P_a \theta(a^i - a_a )$, sets a threshold for atonal at which its production becomes refractive to inhibitory signals.
Finally, the two remaining differential equations for the morphogens \textit{u} and \textit{s}  in Eqs~(\ref{eq:2})~and
~(\ref{eq:3}) have similar functional forms. Here, the theta functions specify the threshold value for atonal in the target cell that is needed for initialization of production.

In this model parameter sets  have to be tuned for the experimentally observed ordered pattern to emerge. Furthermore, as the placement of the R8 cells can vary in cluster size and separation, we use the  biologically most plausible and stable parameter set,  determined by comparison of  numeric simulations to experimental imaging data of developing drosophila ommatidia adopted from \cite{Barkai,Lubensky}, shown in S1~Table. In Fig~\ref{Increasing disorder pattern}(a), we show simulation results for this ideal parameter set at long times after the MF has passed.

\begin{figure}[H]
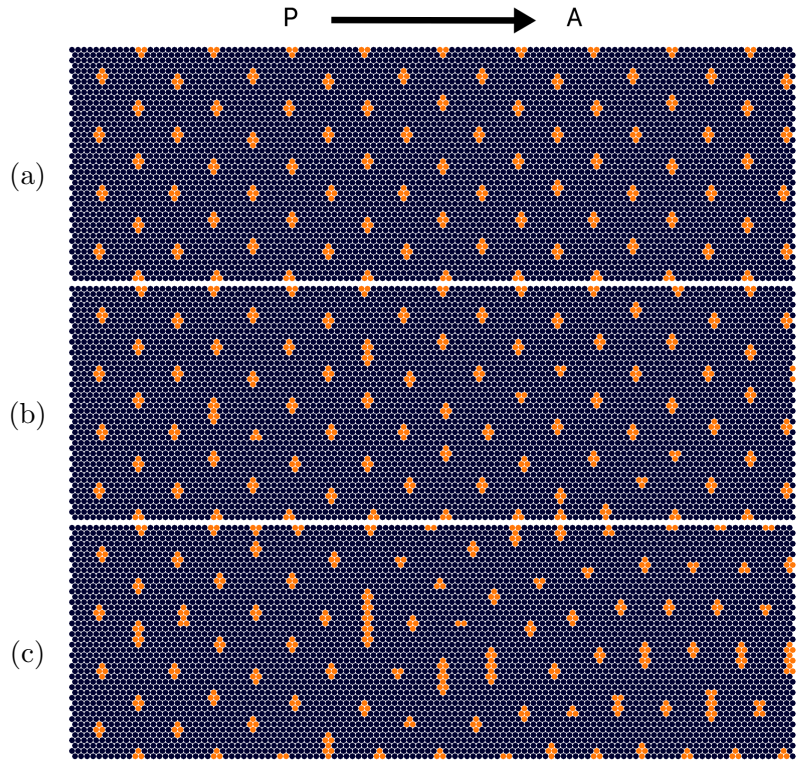

    \centering
    \begin{subfigure}[b]{0.7\textwidth}
     \centering
 \begin{lpic}[l(5mm),r(0mm),t(0mm),b(0mm),grid(10),coords(10),draft,clean]{PtoA(0.4,0.4)}
\end{lpic}
        \label{n0}
    \end{subfigure}
    \begin{subfigure}[b]{0.7\textwidth}
 \begin{lpic}[l(5mm),r(0mm),t(0mm),b(0mm),grid(10),coords(10),draft,clean]{n0(0.4,0.4)}
\lbl[t]{-10,40; (a)}
\end{lpic}
        \label{n0}
    \end{subfigure}

    \begin{subfigure}[b]{0.7\textwidth}
    \begin{lpic}[l(5mm),r(0mm),t(0mm),b(0mm),grid(10),coords(10),draft,clean]{n3_c(0.4,0.4)}
\lbl[t]{-10,40; (b)}
\end{lpic}
        \label{n2}
    \end{subfigure}

    \begin{subfigure}[b]{0.7\textwidth}
    \begin{lpic}[l(5mm),r(0mm),t(0mm),b(0mm),grid(10),coords(10),draft,clean]{n4_c(0.4,0.4)}
\lbl[t]{-10,40; (c)}
\end{lpic}
        \label{n4}
    \end{subfigure}
    \caption{ {\bf Pattern formation simulation results in the Drosophila eye disc, the morphogenetic furrow moves from left to right.}  Cluster positions and sizes of differentiated R8 cells are shown for increasing noise  $\sigma$ in  the diffusivity of the inhibitor D\textsubscript{u}. (a)-(c) are the final patterns for $\sigma / \mu= 0$, $\sigma / \mu= 30\%$, and $\sigma / \mu= 40\%$ respectively. Here, $\mu$  is the mean of the corresponding normal distribution and the value that is used to generate the perfect pattern.}\label{Increasing disorder pattern}
\end{figure}



\subsubsection*{Placement of initial clusters}

 In order for the pattern to propagate, one must place preexisting differentiated cells as an initial condition. Without proper placement, the pattern does not propagate periodically. In this model, a differentiated cell is defined as a cell that expresses a concentration of \textit{a} above a threshold value $a_a$. Prior work has not specified a method for determining the initial differentiated cell placement other than trying all possible configurations.
 
To determine the spacing of the differentiated cells, an isolated differentiated cluster was simulated. After the cell cluster reached a steady state, the shape of the resulting inhibition region was recorded. Then, the spacing between clusters or  single cells was determined by requiring an inhibition region that resembles Fig~\ref{R8visualiation}. Note that the size of a cluster is determined from the properties of the local activator in Eq~\ref{eq:3}, and in its absence single cells are placed instead.

\subsubsection*{Introducing Noise}

Starting from  validated simulations of the noiseless case~\cite{Barkai,Lubensky}, we are now well positioned to study the effects of parametric disorder on  pattern formation. The motivation for this is that even in quasi-identical cells, it is experimentally observed that genes are expressed at appreciably different levels due to stochasticity.  In addition, cells produce morphogens in bursts that randomly diffuse and bind, contributing further to the stochasticity of the system. Here we  investigate how such variations affect the outcome of pattern formation in the Drosophila eye disc. Generalizing the model for the clean system to include the effects of  disorder, each model parameter that appears in the differential equations Eqs~\ref{eq:1}-\ref{eq:4} is  chosen from a normal distribution centered at the mean value that produces the ordered pattern of differentiated R8 clusters in Fig~\ref{Increasing disorder pattern}(a). They are then kept constant during the time evolution. The width and center of each Gaussian distribution is chosen to be the same for all cells. When more noise is introduced, the width of the distribution is increased accordingly. More specifically, in the numerical data discussed below the widths are tuned from 0\% to 60\% of the mean.  As an example, introducing noise in the production rate of \textit{u} corresponds to picking values from the probability distribution:
\begin{equation}  \label{eq:6}
P(P_u) = \frac{1}{\sqrt{2\pi \sigma^2}}\exp{(P_u-\mu)^2/2\sigma^2},
\end{equation}
where $\mu = \langle P_u \rangle$. Introducing noise in the diffusivity of the morphogens is a bit more intricate and is explained in S1~Appendix. The effect of introducing noise in the diffusivity of \textit{u} is presented in Fig~\ref{Increasing disorder pattern}. In this paper, \textsigma{} will refer to the standard deviation of the normal distribution, as shown for \textit{P}\textsubscript{u} in Eq~\ref{eq:6},  and therefore will define the degree of noise. In the unlikely instance that a negative value is drawn from the distribution, the absolute value is considered.

\subsubsection*{Cluster Refinement}

Until now, in the parameter regime deemed biologically plausible, the mathematical model in Eqs~\ref{eq:1}-\ref{eq:4} produces patterns containing clusters. However, the actual developmental process includes an extra step that reduces the clusters to single activated R8 cells, which then become the center of the future ommatidia. This process is associated with the Notch-Delta pathway, whereby the cell that produces the most delta inhibits its nearest neighbors~\cite{Baker}. In order to reproduce this refinement process, a single cell for each cluster is kept as the R8 cell. The cell is chosen to be the most central cell in the most populated row in each cluster,  an approximation to the non-trivial pathways and models identified in the literature ~\cite{Barkai}. This cluster refinement process is illustrated graphically in Fig~\ref{cluster refinement}. This extra step, of cluster refinement, has been analytically shown to increase robustness of the pattern~\cite{gavish2016two}.
\begin{figure}[H]
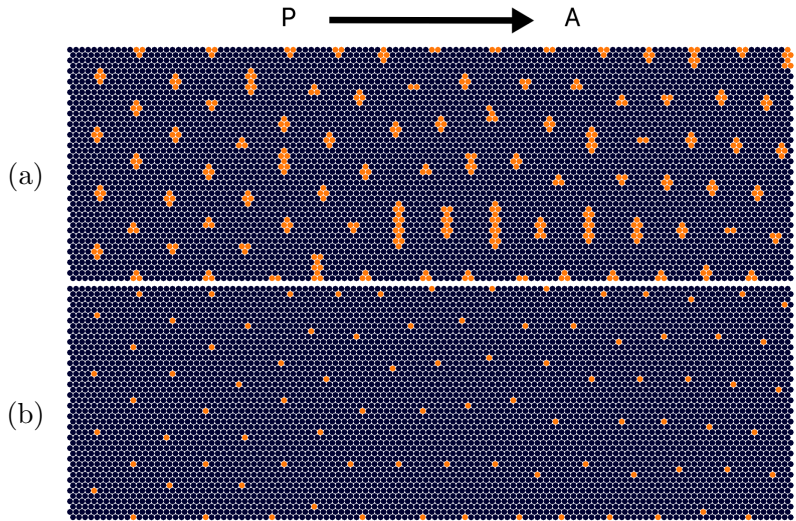

    \centering
    \begin{subfigure}[b]{0.7\textwidth}
     \centering
 \begin{lpic}[l(5mm),r(0mm),t(0mm),b(0mm),grid(10),coords(10),draft,clean]{PtoA(0.4,0.4)}
\end{lpic}
        \label{n0}
    \end{subfigure}
    \begin{subfigure}[b]{0.7\textwidth}
    \begin{lpic}[l(5mm),r(0mm),t(0mm),b(0mm),grid(10),coords(10),draft,clean]{cr0(0.4,0.4)}
\lbl[t]{-10,40; (a)}
\end{lpic}
    \end{subfigure}

    \begin{subfigure}[b]{0.7\textwidth}
    \begin{lpic}[l(5mm),r(0mm),t(0mm),b(0mm),grid(10),coords(10),draft,clean]{cr1(0.4,0.4)}
\lbl[t]{-10,40; (b)}
\end{lpic}
    \end{subfigure}
    \caption{ {\bf Visualization of the cluster refinement step.} The code takes the R8 cell clusters, as shown in (a), as an input and determines the single R8 cell that will become the center of the future ommatidium, as shown in (b). This example is shown for a noise level of $\sigma=40\%$\textmu{}, applied to the diffusivity of u, $D_u$. }\label{cluster refinement}
\end{figure}

\subsection*{Numerical Evaluation}

Numerical forward integration on a 120$\times$44 hexagonal lattice with temporal step size $10^{-2}$ was used to evaluate the spatio-temporal evolution of the coupled differential equations (1)-(4). To obtain sufficient data for statistical analysis, the code was parallelized and run on USC's high-performance supercomputer cluster center. Each computation was assigned to multiple processors at once, running for around 10 hours on each node. 

To obtain reliable error bars, at least 50 realizations of each noise level were simulated. The total eye disc size width was chosen such to eliminate edge effects, allowing sufficient space for clusters to form on the edges. Also, a relatively long eye disk length of 120 lattice sites was used is to allow  investigation of error propagation of clusters as a function of position from posterior (P) to anterior (A). The simulations were terminated once the morphogenetic furrow had passed throughout all of the eye disc.

\subsection*{Quantitative Measures of Structural Order}

While previous quantitative studies have concentrated on characterizing the disorder of the point patterns of experiments and creating minimal underlying models,~\cite{Jiao,Kim} in this work we focus on  the robustness of the mathematical model and its structural characteristics in the presence of stochasticity. This analysis will provide insight into pathways with which developmental systems either cope with or succumb to stochasticity beyond a threshold, ultimately leading to malformation.  

We wish to analyze the emerging point pattern using appropriate measures of structural order, including translational, bond orientation and variance of nearest neighbor distributions. Details of how these measures are designed and applied are discussed in this section. 

Traditional measures for determining structural order have been developed in solid state physics ~\cite{kittel,Egami,TruskettT,Halperin1978theory}. However, in the context of the spatio-temporal formation of patterns in biological developmental models there are additional issues to address. Foremost, the degree of structural order in the final pattern is  generally not homogeneous in the presence of stochasticity. While, per initial conditions, the posterior point pattern  starts off close to an ordered state, it may result in a strongly disordered anterior region at later stages of development, depending on the degree of stochasticity. This type of disorder is correlated between rows because of the manifestly Markovian mechanism that produces the patterns, i.e. the points on a given row are specified based on the geometry of the points previous row \cite{markov1954theory}.  

This observation deems traditional solid state measures used to evaluate the structure of homogeneous crystals insufficient~\cite{Egami}. Furthermore, the type of correlated disorder exhibited in the point patterns of the eye disc is unlike  random thermal displacements found in atoms or infrequent impurities. Instead, such correlated disorder causes Bragg's law to become inapplicable, since it is based on the assumption that the underlying unperturbed lattice is periodic~\cite{kittel}. Specifically, the model system is far away from the two instances where a corrected Bragg's law for imperfect lattices could still be applied. The Debye-Waller approximation~\cite{debye1913interferenz,waller1923frage} applies only for uncorrelated deviations from a perfectly periodic lattice, and the available corrections for correlated deviations~\cite{lindenmeyerParacrystal} apply only in the limit of small deviations and short-ranged Gaussian correlations. In order to study the structure of amorphous solids and general highly disordered materials, as in the case of developmental models, it is therefore necessary to resort to the pair distribution function as a local measure of structural order~\cite{Egami}.

The definition of the  pair distribution function is given by
\begin{equation} g({\bf r})=\frac{1}{N} \sum_{j=1}^N \delta({\bf r}- {\bf r_j}).\end{equation}
In the context of developmental models,  this function  describes the probability of finding an activated  cell at position r, given that another activated  cell is located at position $r_j$. The pair distribution when treated as function of the scalar distance, as in the case of isotropic patterns~\cite{Egami}, is denoted by $g_2(r)$ and is called the radial distribution function.

In the following analysis, we primarily use the radial distribution function. Practically, in numerical simulations of eye disk development one can calculate the radial pair distribution function only for small distances. Each simulation of the pattern only contains around 40 activated cells. Moreover, the calculation of $g_2(r)$ is limited by the open boundary conditions, because the application of periodic boundary conditions to a single realization and the combination of multiple realizations is unphysical. Therefore, here we focus on the radial distribution function for small radii, containing only nearest neighbor activated cells.

Furthermore, we analyze a scalar measure of the spatial order of the patterns, the translational order parameter defined as \begin{equation} \label{eq:7} T=\frac{\int_{0}^{\eta_c}|1-g_2(r)|dr}{\eta_c},\end{equation}  \cite{TruskettT} where \texteta\textsubscript{c} is  a cutoff limited by the simulation size. This order parameter can be interpreted as the Kolmogorov probability distance~\cite{rachev2013methods}  between the radial distribution function of the target pattern and the Poisson random point process. (The radial distribution function of a Poisson random point process is the uniform distribution, which is equal to 1.) The translational order parameter, \textit{T}, is a general order metric used to describe systems independently of the underlying crystal structure. In this setting, we  examine how fast the pattern deteriorates compared to the perfect pattern, so we replace the uniform probability distribution with that of the perfect pattern. In addition, we set the cutoff \texteta\textsubscript{c} equal to the furthest  nearest neighbor distance. 

Finally, we study the bond angle order parameter. Contrary to the translational measure, this order parameter evaluates the spatial orientation of vectors connecting the nearest neighbors of all points. It is defined by 
\begin{equation}  \label{eq:8} q_6=|\frac{1}{N}\sum_{j=1}^N\sum_{k=1}^{N_n} \exp{(6i\theta_{jk})} | ,\end{equation}
\noindent\cite{Halperin1978theory}, taking a value of 1 for a perfect hexagonal point pattern and 0 for a completely random pattern. Here $N_n$ is the total number of nearest neighbors in the point pattern, the  j's sum over all lattice points, and the k's sum over all nearest neighbors of a given reference point. Lastly, $\theta_{jk}$ is the angle of the vector connecting each point with its nearest neighbor with respect to a fixed axis.

\subsubsection*{Probability Distribution Functions of Nearest-Neighbor Distances and Angles}
To quantitatively determine the degree of disorder,  post-processing code is used to record the activated R8 cell positions in the final patterns and generate the probability distributions of nearest-neighbor angles and distances. For every R8 cell, all the neighboring distances are calculated. Then, the nearest neighbors of a cluster are identified as all R8 cells within 1 to 1.5 times the nearest R8 cell distance. 

Next, nearest-neighbor distances and angles of each cluster are calculated and a filter is applied to correct for boundary effects for the nearest neighbor angles. Combining  this information for all random realizations, the probability distribution functions of nearest-neighbor distances and angles are determined. Since the cells in the simulation are positioned on a hexagonal lattice, the nearest-neighbor distances and angles  take discrete values. As a consequence, the probability distributions are also discrete. They are shown in Fig~\ref{No noise histograms} for the ordered case. These types of probability distribution functions are used to quantify the order of the R8 cell point patterns by calculating the variance and the distance between probability distributions. The same approach can be taken when interpreting experimental data obtained from imaging. 

\begin{figure}[H]
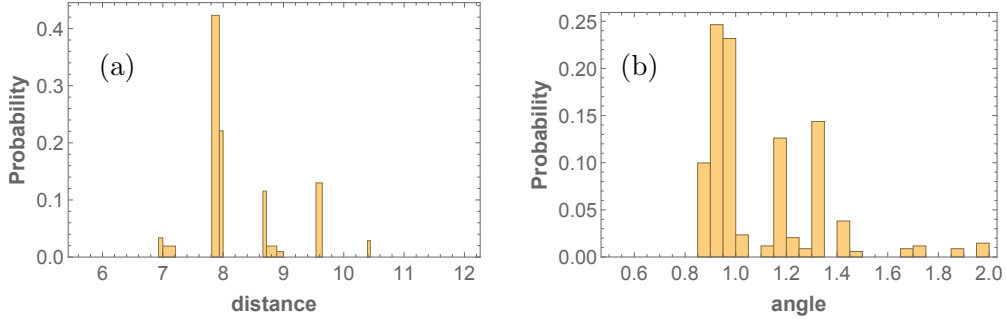

    \centering
    \begin{subfigure}[b]{0.4\textwidth}
    \begin{lpic}[l(0mm),r(0mm),t(0mm),b(0mm),grid(10),coords(10),draft,clean]{hdistanceall0(0.5,0.5)}
\lbl[t]{30,70; (a)}
\end{lpic}
    \end{subfigure}
    \qquad
    \begin{subfigure}[b]{0.4\textwidth}
     \begin{lpic}[l(0mm),r(0mm),t(0mm),b(0mm),grid(10),coords(10),draft,clean]{hangleall0(0.5,0.5)}
\lbl[t]{30,70; (b)}
\end{lpic}
    \end{subfigure}
    \caption{ {\bf  Nearest neighbor (a) distance and (b) angle probability distributions of R8 cell point pattern in the Drosophila eye disc in the absence of stochasticity.} Distances are naturally binned by the available sites on the underlying triangular lattice, whereas angular positions are collected in bins of 0.05 radians.     }\label{No noise histograms}
\end{figure}

\subsubsection*{Variance as a Measure of Order}
The nearest-neighbor angles and distances can be thought of as samples of an underlying distribution. If this distribution has non-vanishing higher-order moments, it is not trivial to produce a measure of disorder. However, in the simple case where the distribution can be  approximated by a Gaussian, its variance, can be used as a reliable measure of the spread of the distribution. The variance used in this context is the sample variance of the nearest neighbor distances and angles. It is calculated from N values,  x\textsubscript{i}, using 
\begin{equation}  \label{eq:9} s^2=\frac{1}{N-1}\sum_{i=1}^{N}{(\overline{x}- x_i)^2}, \end{equation}  
where $\overline{x}$ is the sample mean. This variance can be used to quantify the noise level of the pattern. Since Gaussian distributions have only  non-zero first and second moments, checking whether the higher moments of the histograms vanish with increasing number of realizations can be used to verify that approximating the distribution as Gaussian is reasonable. 

\subsubsection*{Probability Distribution Distance as a Measure of Order: Fidelity and Kolmogorov Distance}

Here we introduce two new and useful measures of order  which best address the needs of this study:  fidelity and  Kolmogorov distance. Both of these rely on the concept of distance between probability distributions, and are generalizations of the translational order parameter $T$ defined in Eq~\ref{eq:7}.  To define these measures of order, we use the zero noise case probability distribution as a reference and calculate its distance from each of the noisy probability distributions. This choice is intuitive, since we are addressing the question of how disordered is the pattern relative to the perfect pattern.

The fidelity of two discrete probability distributions $p(x_i)$ and $q(x_i)$ is defined as 
\begin{equation}  \label{eq:10}F(p(x_i),q(x_i))=\sum_{k} \sqrt{ p(x_k)q(x_k)} ,\end{equation}
where the sum runs over all the bins of the discrete distributions \cite{rachev2013methods}.  The fidelity between two distribution falls into the range  $0\le F(p(x_i),q(x_i)) \le1$, where $1$ is attained only when $p(x_i)=q(x_i) \: \forall \: x_i$. The Kolmogorov distance, in this context, is defined as \begin{equation} \label{eq:11}D(p(x_i),q(x_i))=\frac{1}{2}\sum_{k} |p(x_k)-q(x_k)|, \end{equation} 
where the sum again extends over all the bins of the discrete distributions \cite{rachev2013methods}. Akin to the fidelity, the Kolmogorov distance is bounded by zero and one. This  measure represents the maximum deviation of the two probability distribution functions given that a collection of events occur.

\section*{Results and Interpretation}

\subsection*{Threshold Response}

To quantify  disorder in the emerging activated R8 cell patterns, the nearest neighbor distance and angle probably distribution functions were computed for various noise levels in the diffusion coefficient $D_u$. (Later we will discuss the effect of stochasticity on other model parameters.) Their  histograms are shown in  Fig~\ref{Increasing disorder histograms}. The binning used for the distance histograms reflects the discreteness of the underlying triangular lattice on which the cell centers are placed. For the distance diagrams, all possible nearest neighbor distances  on the triangular lattice are used as bins. For the angle histograms, since there are many more possibilities, a constant bin size of 0.05 radians was chosen in order to appropriately resolve the distribution. 
%
%
\begin{figure}[H]
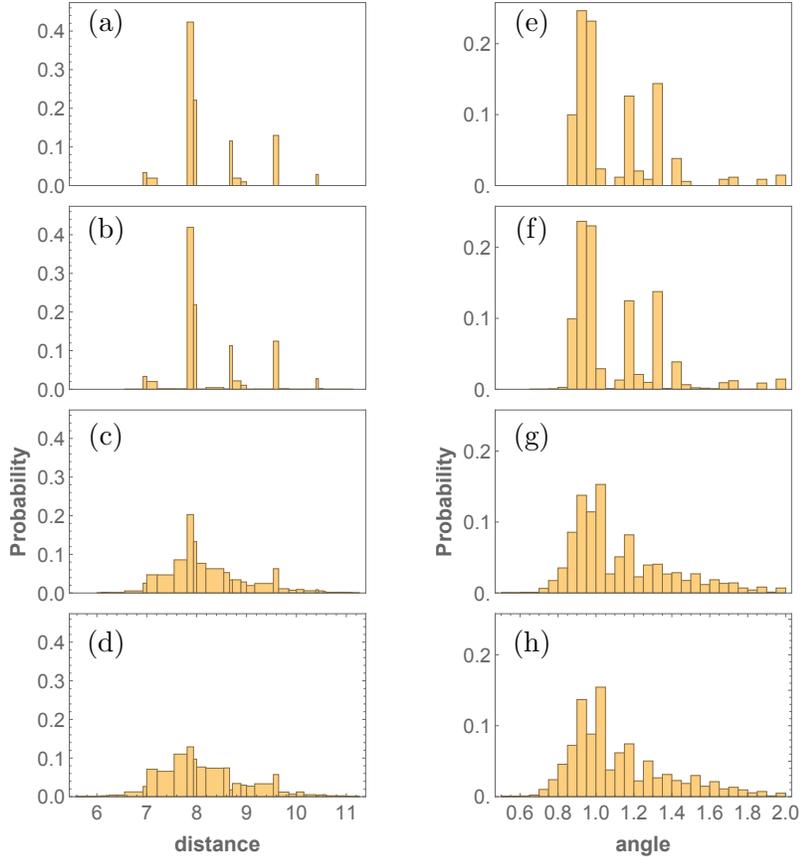

    \centering
    \begin{subfigure}[b]{0.49\textwidth}
     \begin{lpic}[l(0mm),r(0mm),t(0mm),b(0mm),grid(11),coords(11),draft,clean]{verticalhist_alld(0.5,0.5)}
\lbl[t]{44,231; (a)}
\lbl[t]{44,176; (b)}
\lbl[t]{44,121; (c)}
\lbl[t]{44,66; (d)}
\end{lpic}
    \end{subfigure}
    \hspace{-2cm}
    \begin{subfigure}[b]{0.49\textwidth}
    \begin{lpic}[l(0mm),r(0mm),t(0mm),b(0mm),grid(11),coords(11),draft,clean]{verticalhist_alla(0.5,0.5)}
\lbl[t]{44,231; (e)}
\lbl[t]{44,176; (f)}
\lbl[t]{44,121; (g)}
\lbl[t]{44,66; (h)}
\end{lpic}
    \end{subfigure}
    \caption{{\bf Nearest neighbor distance (left panels) and angle (right panels) probability distributions generated from the point-pattern in the Drosophila eye disc R8 cell specification for increasing noise levels.} Stochasticity was introduced in the differential equations by drawing the value of the diffusion coefficient $D_u$ from a normal distribution with mean $\mu $ and  standard deviation $\sigma$. The  nearest neighbor distance histograms in (a) to (d) and nearest neighbor angle histograms in (e) to (h)  correspond Gaussian stochasticity with standard deviations $\sigma / \mu=0\%,20\%,40\%,60\%$. While the first two histograms  are almost identical, as noise in $D_u$ increases, the peaks get smeared out and the histograms appear wider.}\label{Increasing disorder histograms}
\end{figure}
The angle histograms, unlike the distance histograms, are skewed towards larger angles.  This asymmetry makes the analysis of the distributions more complicated, as  in this case the variance, paints an incomplete picture. As discussed below, the probability distance order measure eliminates this issue. The angle histograms are skewed as a direct consequence of the cluster refinement process. As seen in the plots of atonal patterns (Fig. 3), the onset of disorder is signaled by elongated clusters of activated R8 cells. Since these elongated clusters are brought down to a single cell in the cluster refinement step, and they cast large inhibition radii, the number of nearest neighbors found for the elongated clusters is less than the six found in the perfect pattern. A lesser number of neighbors trivially leads to higher angles in this case.  Both histograms, distances and angles, start out sparse for the case without any stochasticity in $D_u$, with empty bins between peaks, indicative of the highly ordered repeating lattice structure produced by the simulations in this case. As stochasticity increases, the resulting point patterns become aperiodic, leading to denser histograms.
Furthermore, the histograms  become wider with increasing stochasticity. This is captured in the variance, shown in Fig~\ref{Variance order parameter}.  

\begin{figure}[H]
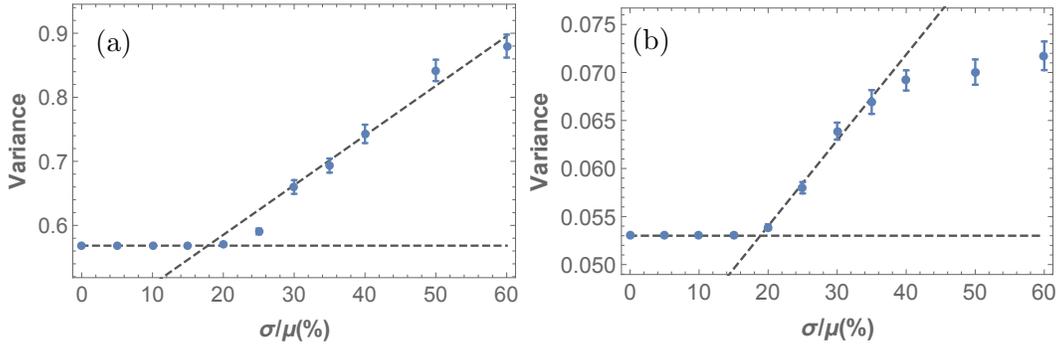

    \centering
    \begin{subfigure}[b]{0.45\textwidth}
    \begin{lpic}[l(0mm),r(0mm),t(0mm),b(0mm),grid(10),coords(10),draft,clean]{v1(,45mm)}
\lbl[t]{27,77; (a)}
\end{lpic}
    \end{subfigure}
    \begin{subfigure}[b]{0.467\textwidth}
     \begin{lpic}[l(0mm),r(0mm),t(0mm),b(0mm),grid(10),coords(10),draft,clean]{v2(,44.5mm)}
\lbl[t]{30,75; (b)}
\end{lpic}
    \end{subfigure}
	\caption{ {\bf Disorder in the R8 cell point pattern of the developing  Drosophila eye disc as a function of stochasticity.}
Here, the variance of the distributions of (a) nearest neighbor distances and (b) nearest neighbor angles is used to quantify the amount of disorder. 
As in  the previous figure, stochasticity is introduced in the diffusivity $D_u$, and the x-axis represents the noise level in terms of the standard deviation of the underlying Gaussian distribution from which this parameter is drawn. As the noise level in the simulation is increased beyond a threshold, the variance grows. The slight irregularity of the angle plot between $20\%$ and $30 \%$ is a result of the skewness of the nearest neighbor angle distributions. The angle variance saturates after $40\%$. }\label{Variance order parameter}
\end{figure}

The variance plots in Fig~\ref{Variance order parameter} illustrate the generic nature of  pattern formation in stochastic developmental models. They show that for small levels of stochasticity (here $\sigma / \mu < \sigma_c / \mu=20 \% $) the ordered pattern remains basically unaffected. However, beyond this threshold the variance starts increasing linearly, both in the angle and distance histograms. This threshold response is the central finding of this study. As discussed below, this is a generic phenomenon, largely independent of which model parameters experience stochasticity and which measures of order are used to assess the system response. It implies that these stochastic models capture biological resilience against stochastic variation up to a certain point, resulting in a regime of deterministic outcomes in spite of stochastic input. 

Let us now discuss how this type of threshold behavior is also picked up by the more general measures of order given by Eqs~\ref{eq:10},\ref{eq:11}. The results of applying these probability distance  measures is shown in Fig~\ref{Independence of measure}. Note that  saturation observed in the nearest angle distribution variance plot is apparent in both cases, and the responses of both observables (distances and angles)  to stochasticity exhibit a universal sigmoidal functional form, implying threshold behavior. These measures reveal the same interesting quality as the variance: the pattern order exhibits an initial resistance to weak input stochasticity that eventually gives way to structural malformation at larger noise levels. The sigmoidal functional form confirms that the mechanism does exhibit robustness for $\sigma < \sigma_c$. The fact that this is observed in  the Kolmogorov measure, the  fidelity, the bond orientation and variance measures supports the notion that such threshold response to stochasticity is a generic feature (see supporting information, S1~Fig, for bond orientation sigmoidal response).

\begin{figure}[H]
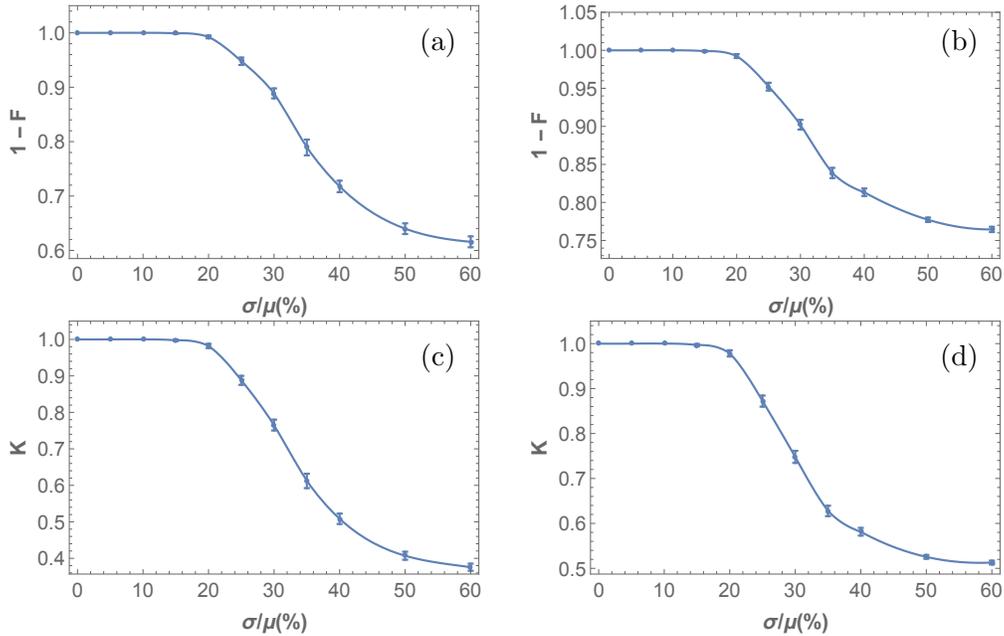

    \centering
    \begin{subfigure}[b]{0.45\textwidth}
     \begin{lpic}[l(0mm),r(0mm),t(0mm),b(0mm),grid(11),coords(11),draft,clean]{o1(0.5,0.5)}
\lbl[t]{115,77; (a)}
\end{lpic}
    \end{subfigure}
    \begin{subfigure}[b]{0.45\textwidth}
    \begin{lpic}[l(0mm),r(0mm),t(0mm),b(0mm),grid(11),coords(11),draft,clean]{o2(0.5,0.5)}
\lbl[t]{115,77; (b)}
\end{lpic}
    \end{subfigure}
    
    \begin{subfigure}[b]{0.45\textwidth}
    \begin{lpic}[l(0mm),r(0mm),t(0mm),b(0mm),grid(11),coords(11),draft,clean]{k1(0.5,0.5)}
\lbl[t]{115,77; (c)}
\end{lpic}
     \end{subfigure}
    \begin{subfigure}[b]{0.45\textwidth}
    \begin{lpic}[l(0mm),r(0mm),t(0mm),b(0mm),grid(11),coords(11),draft,clean]{k2(0.5,0.5)}
\lbl[t]{115,77; (d)}
\end{lpic}
    \end{subfigure}
    \caption{ {\bf Probability distance measures applied to analyze response to stochasticity in R8 cell point pattern formation.} (a)-(b) are generated using the fidelity, F, and (c)-(d) using the Kolmogorov distance, K, for  nearest neighbor distances and angles respectively. Stochasticity was introduced to the model through drawing the parameter $D_u$ in the differential equations Eqs~\ref{eq:1}-\ref{eq:4} from a normal distribution with varying standard deviation, $\sigma$, and mean value, \textmu .  The simulation results were compiled, and nearest R8 cell neighbor angle and distance distributions were obtained. Kolmogorov and fidelity of the resulting distributions where computed with reference to the perfect pattern case. Here, a value of 1 corresponds to a perfectly ordered pattern, and a value of 0 corresponds to a completely irregular pattern. This figure illustrates that the functional form is  independent of the measure used.}\label{Independence of measure}
\end{figure}

    
Next, we turn to the question of universality with respect to how stochasticity is introduced. We observe that stochasticity in other model parameters also produces threshold behavior, indicating that the observed threshold response is a generic feature of this class of evolutionary models. Specifically, in Fig~\ref{Independent of parameter} we show the response to increasing stochasticity levels in the production rate of atonal $P_a$, the production rate due to the activator \textit{S}, and the production rate caused by the morphogenetic furrow \textit{G}. In all these cases, there is also a regime $\sigma < \sigma_c$ where the ordered pattern remains intact. 

\begin{figure}[H]
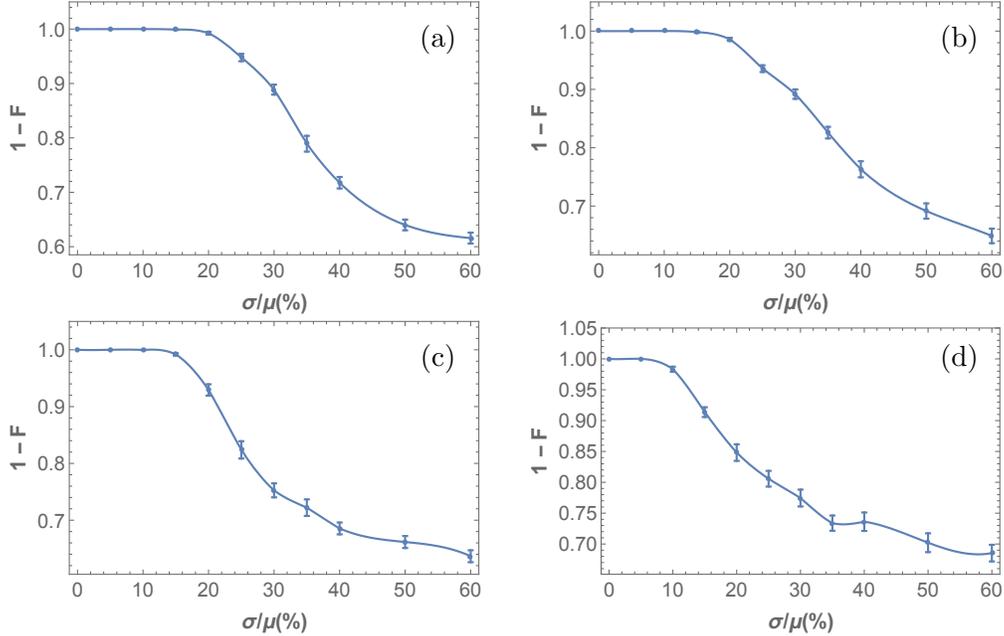

    \centering
    \begin{subfigure}[b]{0.45\textwidth}
     \begin{lpic}[l(0mm),r(0mm),t(0mm),b(0mm),grid(11),coords(11),draft,clean]{o1(0.5,0.5)}
\lbl[t]{115,77; (a)}
\end{lpic}
    \end{subfigure}
    \begin{subfigure}[b]{0.45\textwidth}
    \begin{lpic}[l(0mm),r(0mm),t(0mm),b(0mm),grid(11),coords(11),draft,clean]{o1_Pa(0.5,0.5)}
\lbl[t]{115,77; (b)}
\end{lpic}
    \end{subfigure}
    
    \begin{subfigure}[b]{0.45\textwidth}
    \begin{lpic}[l(0mm),r(0mm),t(0mm),b(0mm),grid(11),coords(11),draft,clean]{o1_S(0.5,0.5)}
\lbl[t]{115,77; (c)}
\end{lpic}
     \end{subfigure}
    \begin{subfigure}[b]{0.45\textwidth}
    \begin{lpic}[l(0mm),r(0mm),t(0mm),b(0mm),grid(11),coords(11),draft,clean]{o1_G(0.5,0.5)}
\lbl[t]{115,77; (d)}
\end{lpic}
    \end{subfigure}
    \caption{ {\bf Universal threshold response to  stochasticity in different model parameters.} 
    In each sub-plot, stochasticity is introduced in a single parameter of the underlying differential equations: (a) $D_u$, (b) $P_a$,  (c) $S$, and  (d) $G$. Introduction of parametric noise in all of the parameters, produces a sigmoid response. }\label{Independent of parameter}
\end{figure}
Next let us address the origin of the observed resilience of this model towards weak noise with $\sigma < \sigma_c$, signified by the initial plateaus observed in 
 Fig~\ref{Independent of parameter}. The main reason for these wide regimes, exhibiting robustness with respect to parameter stochasticity, is the presence of  Scabrous as a diffusible activator \cite{Barkai}. While this ingredient is not needed for the pattern to propagate, without this element even low stochasticity levels of $D_u$ cause it to deteriorate, making the model of pattern formation biologically implausible~\cite{Barkai}. This effect is quantitatively demonstrated in Fig~\ref{With and without s}, which compares the response to stochasticity with and without Scabrous as a diffusible activator. In the absence of Scabrous, with the same parameter set and with optimized initial conditions, there is an immediate deterioration of the patterned order upon introduction of infinitesimal noise. In contrast, the presence  of  Scabrous changes this functional relationship to the observed sigmoid response,  thus introducing  a resilience scale to the pattern formation mechanism set by the Scabrous production rate $P_s$ and the activation threshold $s_1$. 

\begin{figure}[H]
    \centering
    \begin{subfigure}[b]{0.85\textwidth}
        \includegraphics[width=\textwidth]{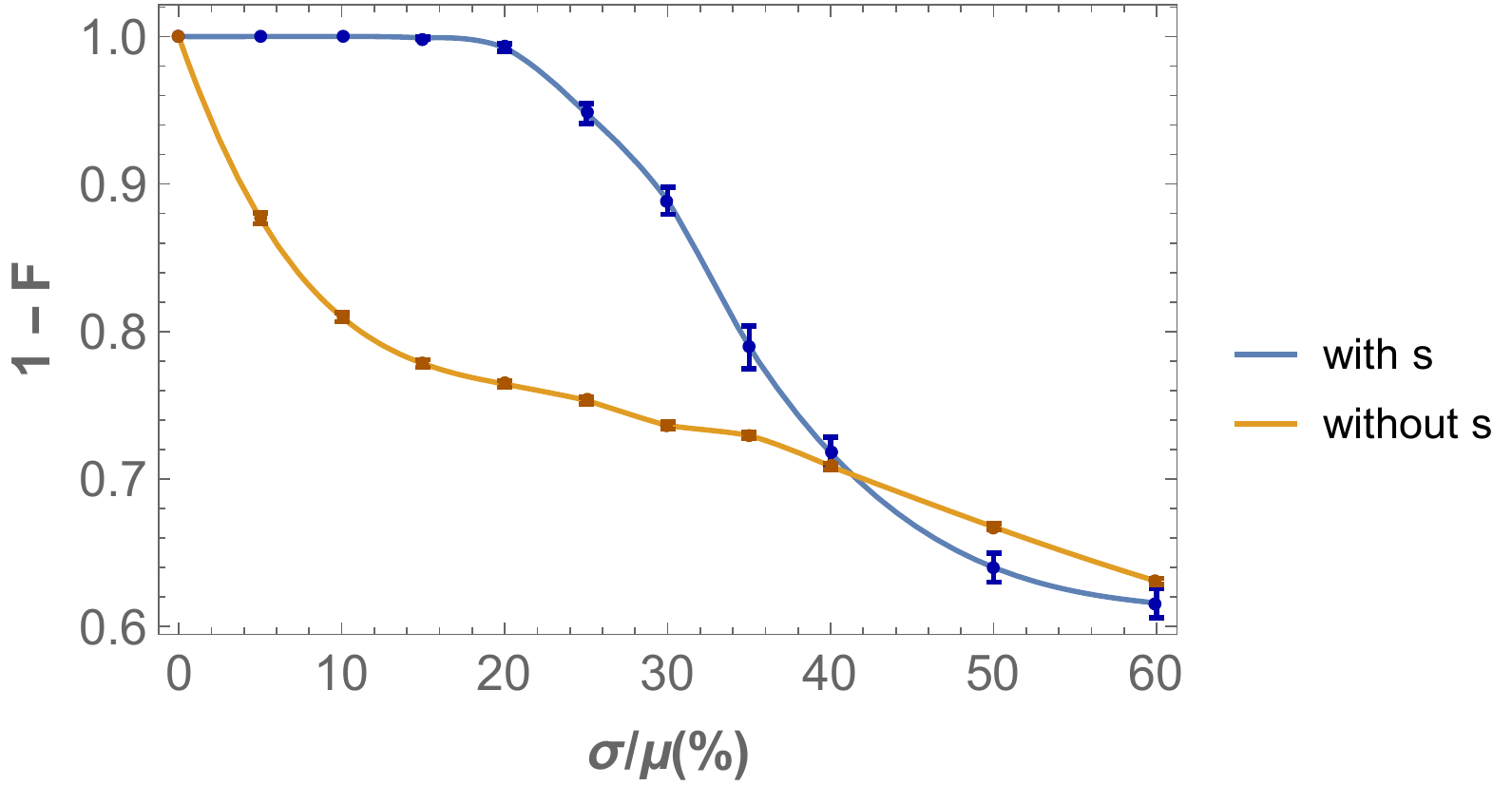}
        \label{DFEAP}
    \end{subfigure}
   
    \caption{ {\bf Robustness of Drosophila eye disc R8 cell pattern order versus stochasticity, with and without the diffusible activator Scabrous in the model.} The analysis is applied in the same setting as Fig~\ref{Independence of measure}, i.e. stochasticity is introduced in the parameter  $D_u$. In the absence of the diffusible activator $s$, there is a much increased sensitivity to stochasticity for infinitesimal non-vanishing $\sigma$. 
    }\label{With and without s}
\end{figure}

In addition to the robustness introduced by  Scabrous, the threshold response of the order of the pattern to stochasticity in the various parameters shown in Fig~\ref{Independent of parameter} is a consequence of three further elements: (i) the spatial discreteness of the underlying lattice, (ii) the threshold activated response of cells to morphogen levels, and (iii) propagation of avalanches from anterior to posterior. As we now discuss, the robustness for low stochasticity levels is a consequence of the lattice discreteness and the threshold response of the cells to morphogen concentrations, whereas the sharp decay is a consequence of propagation of avalanches.

To understand how the discreteness of the underlying lattice contributes to robustness for low noise levels, consider the effect of altering the diffusivity of a morphogen. Since the local morphogen concentration drops off at distances on the order of a few cells, and the cells lie on a discrete triangular lattice, a small change in the diffusivity radius will not result in any difference unless it leads to the inclusion or exclusion of additional cells. A continuous change in the radius of inhibition will exclude or include new cells when it increases by a factor comparable to the lattice spacing, thus creating a threshold response. Alternatively, consider the effect of the morphogenetic furrow, mathematically denoted by Eq~\ref{eq:4}, which propagates with a constant speed through the lattice of cells. Two consecutive cells on along the direction of propagation of the morphogenetic furrow will receive the activation signal with a delay lag, $\delta t$. Only if the production rates, denoted by G in Eg.~\ref{eq:1}, picked from the normal distributions differ above a finite threshold one will result to a misplaced cell. An analytical approach to this  effect and its contribution in error is carried out in \cite{gavish2016two}.

Second, the threshold response of pattern formation to parameter stochasticity is linked to the  a threshold activation of cells in response to morphogen levels. The turn-on and turn-off of morphogens in the differential equations is governed by step functions with threshold values, $a_a$, $a_u$, $u_1$, $h_1$, $s_1$. This has the consequence that the effect of morphogen concentration is binary, i.e. the parameter variation must be sufficient such that the concentration threshold is crossed for activation or inhibition to occur.

Finally, to explain the fast decay of the pattern beyond threshold stochasticity, the propagation of avalanches comes into play, an effect also identified by \cite{Barkai}. Beyond $\sigma_c$,  R8 clusters start to get significantly elongated and misplaced.  The first such elongated R8 cell cluster causes the next row of R8 cell clusters to be misplaced, leading to a cascade onset of pattern disorder. This avalanche effect is quantitatively discussed next, where we analyze the order of the pattern as a function of position. 

\subsection*{Increasing Disorder of the Pattern from Posterior to Anterior}
Here we use the fidelity measure to evaluate error propagation as a function of position on the spatial posterior-to-anterior axis. The mechanism of pattern formation relies delicately on carefully spaced precursor R8 cells that are placed in the posterior region before the growth process described by  the differential equations is initiated. Their positions define the subsequent cluster spacing of activated R8 cells for the entire eye disc. In the presence of stochasticity, this information dissipates during the propagation of the morphogenetic furrow from posterior to anterior, since the ommatidial pattern on a column is defined by the shape of the inhibition signals emitted by the ommatidial row before it. It is thus expected that as the morphogenetic furrow travels from posterior to anterior, errors in the pattern order grow in an avalanche fashion. This effect is observed in the simulations, as shown in Fig~\ref{Functional form of noise P-A}, where the disorder of the pattern clearly increases from posterior to anterior. 

\begin{figure}[H]
    \centering
    
    \begin{subfigure}[b]{0.85\textwidth}
        \includegraphics[width=\textwidth]{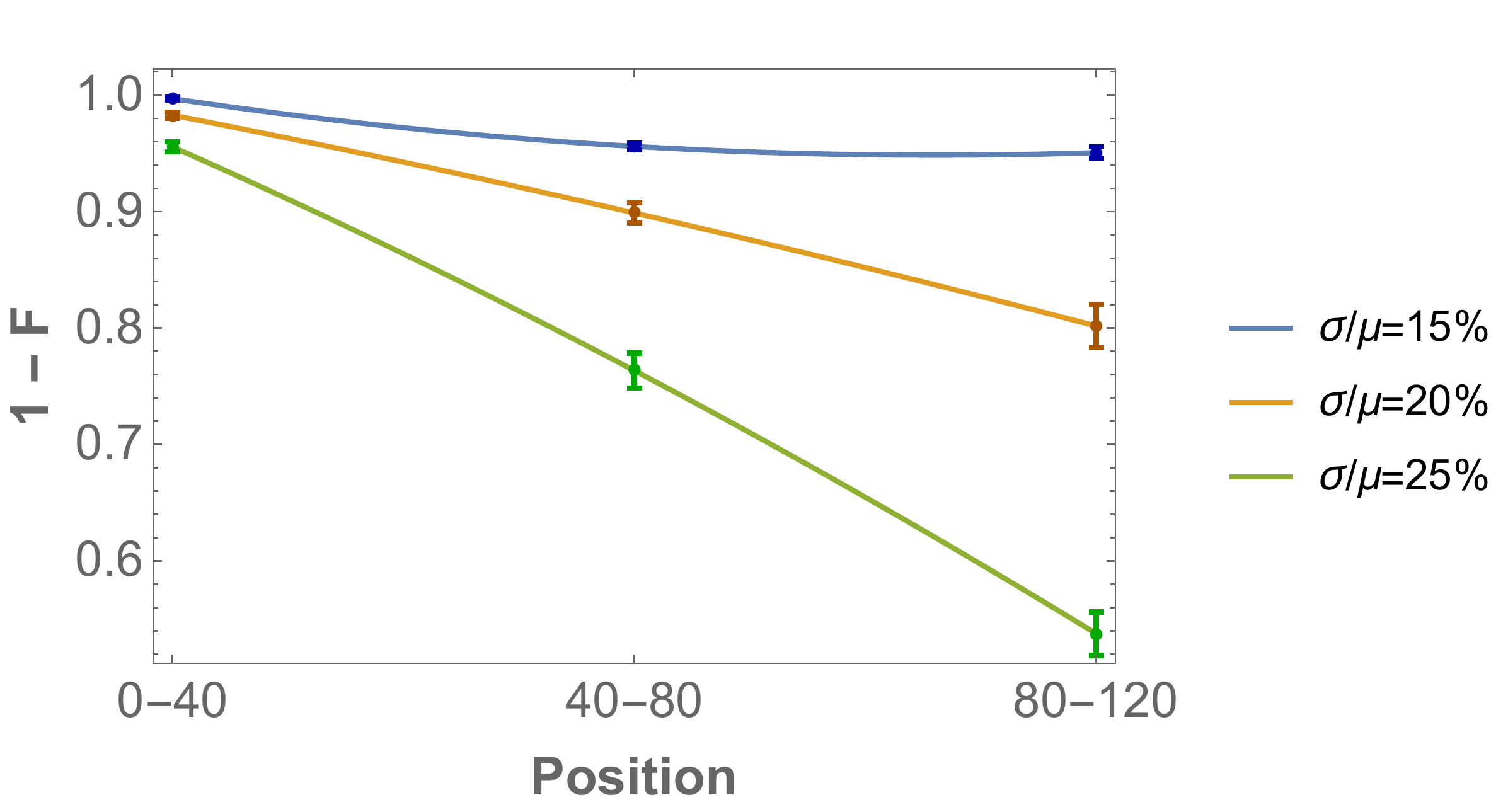}
        \label{DFEAP}
    \end{subfigure}
   
    \caption{ {\bf Decay of R8 pattern order as a function of position on the eye disc} (same parameter set as in Fig. 5). The local order measure is calculated as a function of position, from posterior to anterior. The x-axis refers to regions of the simulated eye disc 0-40,40-80,80-120 respectively, whereas the y-axis refers to the fidelity probability distance measure applied to  nearest neighbor distance distributions. The three colored lines correspond to various stochasticity levels in $D_u$, from $\sigma/\mu = 15\%$ to $\sigma/\mu = 25\%$ in steps of $5\%$. Local order generally decreases from posterior to anterior. Once the threshold of $\sigma_c/\mu \approx 20\%$ is crossed, there is a very sharp spatial decay of the order of the pattern, signaling the underlying avalanche effect. }\label{Functional form of noise P-A}
\end{figure}
Previous studies have not, so far, quantitatively analyzed the consequences of finite system sizes on the order of the pattern. 
To illustrate these finite size effects on the threshold response quantitatively, in Fig~\ref{Effect of size} we plot the order parameter obtained from numerical simulations for different eye disc sizes, illustrating the effect of size on the robustness of the system. Here one clearly observes that the larger eye-disc, the sharper the threshold response to stochasticity.
Since our numerical simulations were performed on lattices much smaller than the actual eye disc, the observed threshold effect can be extrapolated to be much more pronounced for realistic eye disc sizes. 

\begin{figure}[H]
    \centering
    \begin{subfigure}[b]{0.85\textwidth}
        \includegraphics[width=\textwidth]{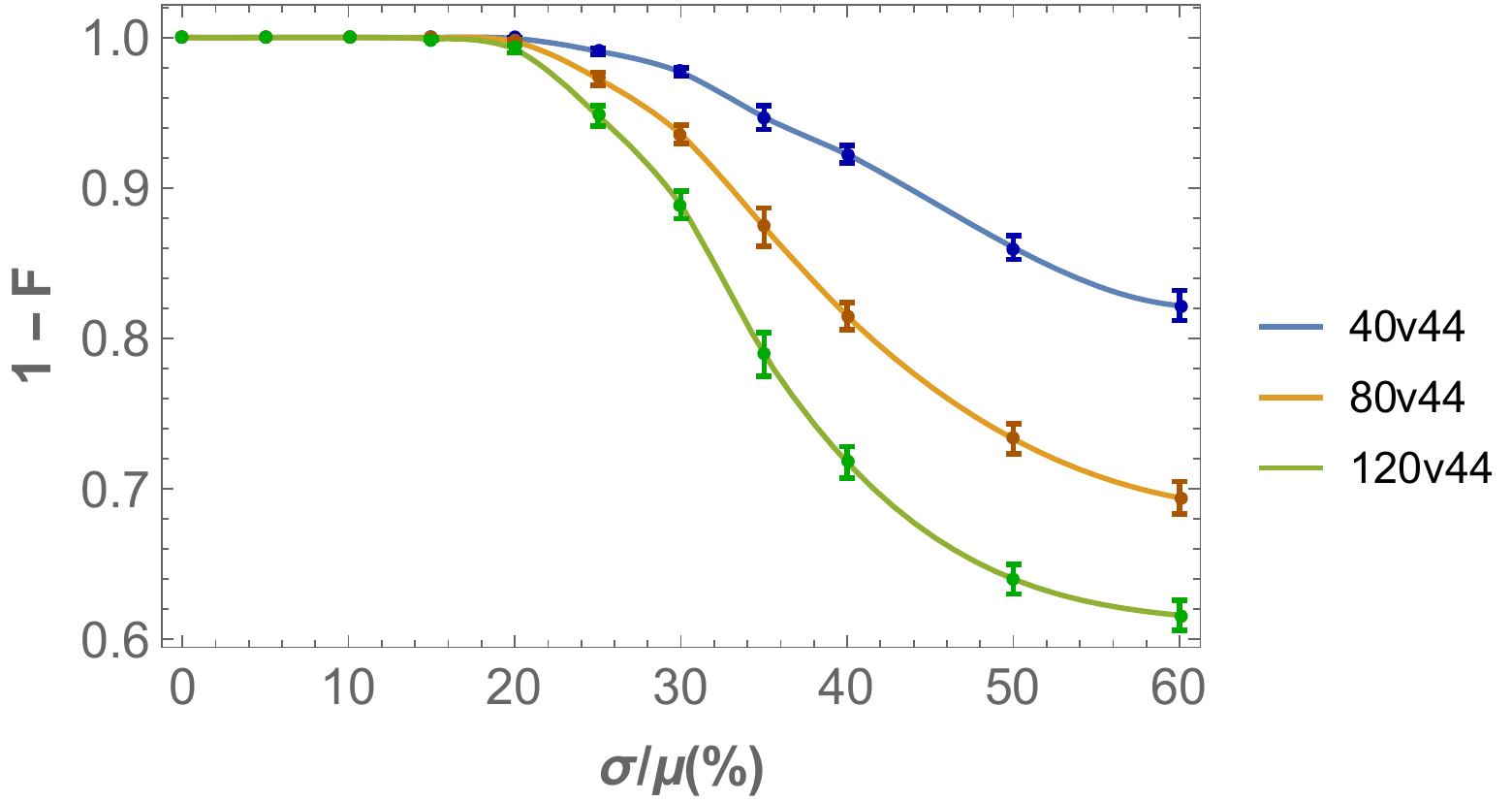}
        \label{DFEAP}
    \end{subfigure}
   
    \caption{{\bf Pattern order as a function of eye disc size} (same parameter sets as in Fig 8). The threshold response is more pronounced as the length of the eye disc in the direction of the morphogenetic furrow is increased. The order parameter is calculated for cell number 40, 80 and 120 while keeping the size perpendicular to the propagation direction of the morphogenetic furrow at 44 cells.  }\label{Effect of size}
\end{figure}

This variation in the degree of dependency of the pattern order measure upon stochasticity levels across different eye disk sizes is analogous to the finite size scaling of order parameters commonly observed in interacting physical many body systems, such as Ising models on lattices. In analogy to phase transitions in the thermodynamic limit of  such models, i.e. for infinite system sizes, here we observe precursor sigmoidal response to increasing stochasticity levels that becomes more pronounced for larger system (eye disk) sizes, ultimately culminating into a true phase transition as $L_x  \rightarrow \infty$. The structure of the underlying system of coupled dynamical equations suggests that this transition belongs to the universality class of directed percolation~\cite{chowdhurystatistical}. While a complete scaling analysis is beyond the scope of this study, it will be provided in a future publication.

\section*{Conclusions}
The  model of coupled  differential equations in Eqs~\ref{eq:1}-\ref{eq:4}, describing eye disc R8 cell differentiation, provides robust developmental mechanisms that lead to  resilience against stochastic perturbations. This robustness can be attributed to the discrete nature of the underlying lattice and the threshold activation of cells to morphogens. In addition, the introduction of Scabrus, as rigorously illustrated in \cite{Barkai}, leads to increased resilience against stochasticity, especially in stochasticity in the diffusivity of the inhibitor. Beyond a critical noise threshold $\sigma_c$, there is an acute loss of order with increasing stochasticity due to avalanches of misplaced differentiated cells. 

This threshold response to noise is an essential characteristic of developmental systems. A lack of resilience would deem biological systems unable to produce complex, nearly perfect structures in an inherently noisy background. This is especially true in eye disc formation with large numbers of ~800 ommatidia. The introduction of Scabrus creates a redundancy that supports the threshold response. The model system studied here exhibits biologically plausible robustness, identifying the sigmoid functional form as the generic response to noise of developmental models. 

The expression of a gene depends on a probabilistic outcome determined by the upstream regulatory sequences that act in cis, and the binding and processivity of other molecules that act in trans. Thus, mutational changes in cis-regulatory elements and trans-regulatory factors can affect transcription levels and transcriptional noise \cite{metzger2016contrasting}. These changes in transcriptional noise reveal some of the evolutionary constraints on cis-regulatory elements~\cite{metzger2015selection}.
The sigmoid functional response  buffers the system against perturbations. This functional form indicates the effect of Cryptic Genetic Variation (CGV).  CGV is largely neutral until it is exposed in certain genetic backgrounds. Robustness in development allows for the accumulation of CGV without any observable changes in phenotype. Once the genetic background changes, for example through mutational perturbation, this previously-neutral bottled-up CGV can be released to produce strikingly different phenotypes~\cite{gibson2004uncovering}. 

Our analysis further illustrates that as the R8 clusters are specified, errors propagate, leading to increased irregularity from posterior to anterior. This rate is  important as it suggests that there is an interplay between how large an eye disc is with how perfectly ordered it can be, with larger eye discs being more likely to accumulate some positional errors. This has interesting implications on limiting eye disc size, a hypothesis that can be investigated experimentally. 

The  structural order measures outlined in this paper can be used to quantify order of patterns in developmental systems containing local point patterns. The application of the radial pair correlation function for  nearest neighbors, as specifically performed in this study, is useful for data sets that have non trivial periodic structure and are small in size. Specifically,  the measure is useful for analysis of experimental data of the eye-disc, since the effect of curvature of the eye-disc is eliminated and restrictive boundary effects are resolved. The point-pattern order measure used here, based on probability distance, is independent of the functional form of the underlying nearest neighbor distributions and boundary conditions, thus making it applicable independent of the form of the point pattern.

\newpage

\section*{Supporting Information}

\paragraph*{S1 Appendix.}
\label{S1 Appendix}
{\bf Exact expression of the Laplace operator and noise.}
\newline \newline
 The discretized form of the Laplace operator, in the triangle lattice arrangement of cells, is given by: $ D_u\Delta u^i = \sum_{<ij>}D_u^{ij}(u_i-u_j)$. Where $D_u^{ij}=D_u^{ji}$ and $<ij>$ denotes all six nearest neighbors, rigorously defined by via a Voronoi diagram of the triangle lattice. Only in the case where there is no noise, $D_u^{ij}=D_u$ and the expression simplifies.

\paragraph*{S1 Table.}
\label{S1_Table}
{\bf Numerical parameters of complete mathematical model.} This parameter set is taken from previous studies, plugging it in the mathematical equations produces the desired pattern in a biologically plausible regime~\cite{Barkai}.

\begin{center}
\begin{tabular}{ |c|c|c|c|c|c|c| } 
 \hline
 $\tau_a=1.2$ & $P_a=3$          & $D_h=600$ & $a_a=0.4 $   & $h_1=1$ & $P_s=10$& $S=2.5$\\ 
 $\tau_h= 100 $ & $P_h=2.4$    & $D_u=0.1 $  & $a_u=1.55 $ & $u_1=2\cdot10^-6$ &$\lambda_s=1$ & $D_s=0.016$\\ 
 $\tau_u=0.015$ & $P_u=10$  & $G=0.9$    & v=0.75  & $\lambda_a=\lambda_h=\lambda_u=1$ & $\tau_s=0.01$ & $s_1=0.01$ \\ 
 \hline
\end{tabular}
\end{center}

\begin{center}
\begin{tabular}{ |c| } 
 \hline
 $c_1=\sqrt[]{(v\tau_h)^2+4D_h}$ \\ 
 \hline
\end{tabular}
\end{center}

\paragraph*{S1 Fig.}
\label{S1_Fig}
{\bf Sigmoid response of bond orientation order parameter.} The bond orientation order parameter also exhibits the sigmoid response to noise. This further supports the fact that the response of the system to noise is physical. To calculate the bond orientation order, the Voronoi diagram method was used to precisely determine the nearest neighbors in the point pattern.

\begin{figure}[H]
    \centering
    \begin{subfigure}[b]{0.85\textwidth}
        \includegraphics[width=\textwidth]{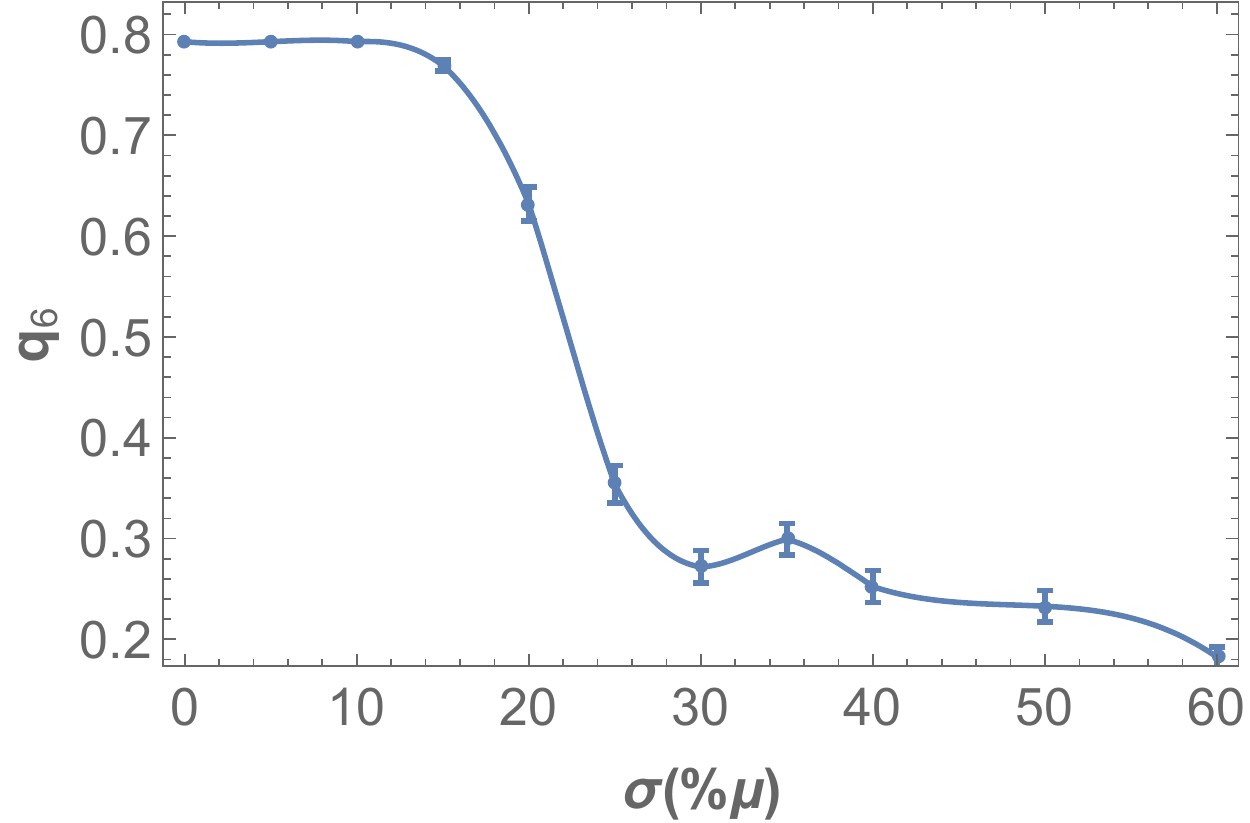}
        \label{ok}
    \end{subfigure}
   \label{bond orientation}
\end{figure}

\section*{Acknowledgements}

We thank Avishai Gavish, first author of \cite{Barkai}, for answering multiple questions regarding the mathematical model adopted in this paper. In addition, Alexander Samsonov, Simon Restreppo and Georgios Styliaris for valuable discussions. Computation for the work described in this paper was supported by the University of Southern California's Center for High-Performance Computing (hpc.usc.edu).
\bibliographystyle{vancouver}
\bibliography{Bibliography}

\end{document}